\def\year{2024}
\documentclass[letterpaper]{article} 
\usepackage{xcolor}
\usepackage{soul}
\usepackage{comment}
\usepackage{booktabs}
\usepackage{aaai24}  
\usepackage{times}  
\usepackage{helvet}  
\usepackage{courier}  
\usepackage[hyphens]{url}  
\usepackage{graphicx} 
\usepackage{natbib}  
\usepackage{caption} 
\usepackage{algorithm}
\usepackage{algorithmic}
\usepackage{tabularx}
\usepackage{multirow}

\usepackage{newfloat}
\usepackage{listings}
\DeclareCaptionStyle{ruled}{labelfont=normalfont,labelsep=colon,strut=off} 
\floatstyle{ruled}
\newfloat{listing}{tb}{lst}{}
\floatname{listing}{Listing}
\newcommand\anna[1]{\textcolor{blue}{AK: #1 }}

\title{Do Responsible AI Artifacts Advance Stakeholder Goals? \\Four Key Barriers Perceived by Legal and Civil Stakeholders} 

\author{
    Anna Kawakami\textsuperscript{\rm 1}\thanks{The author completed this work while at Microsoft Research.},
    Daricia Wilkinson\textsuperscript{\rm 2},
    Alexandra Chouldechova\textsuperscript{\rm 3} \\
}

\affiliations {
    \textsuperscript{\rm 1}Carnegie Mellon University,
     \textsuperscript{\rm 2}Arizona State University,
     \textsuperscript{\rm 3}Microsoft Research\\
     akawakam@andrew.cmu.edu, daricia.wilkinson@asu.edu, alexandrac@microsoft.com
}

\usepackage{bibentry}
\begin{document}

\maketitle


\begin{abstract}

The responsible AI (RAI) community has introduced numerous processes and artifacts---such as Model Cards, Transparency Notes, and Data Cards---to facilitate transparency and support the governance of AI systems.  While originally designed to scaffold and document AI development processes in technology companies, these artifacts are becoming central components of regulatory compliance under recent regulations such as the EU AI Act. 
Much of the existing literature has focussed on the design of new RAI artifacts or their use by practitioners within technology companies.  However, as RAI artifacts begin to play key roles in enabling external oversight, it becomes critical to understand how stakeholders---particularly stakeholders situated \textit{outside} of technology companies who govern and audit industry AI deployments---perceive the efficacy of RAI artifacts.  In this study, we conduct semi-structured interviews and design activities with 19 government, legal, and civil society stakeholders who inform policy and advocacy around responsible AI efforts. While participants believe that RAI artifacts are a valuable contribution to the broader AI governance ecosystem, many have concerns around their potential unintended and longer-term impacts on actors outside of technology companies (e.g., downstream end-users, policymakers, civil society stakeholders). We organized these beliefs into four barriers that help explain how RAI artifacts may (inadvertently) reconfigure power relations across civil society, government, and industry, impeding civil society and legal stakeholders' ability to protect downstream end-users from potential AI harms. Participants envision how structural changes, along with changes in how RAI artifacts are designed, used, and governed, could help re-direct the role of artifacts to support more collaborative and proactive external oversight of AI systems. 
We discuss research and policy implications for RAI artifacts. 


\end{abstract}

\maketitle



\section{Introduction}\label{Introduction}
Technology companies are now developing and deploying artificial intelligence (AI) at unprecedented speeds and scales \cite{muthukrishnan2020brief}. In
turn, the responsible AI (RAI) community has rapidly innovated new RAI artifacts—processes, tools, and other resources designed to support the ethical creation and use of AI ~\cite{reisman2018algorithmic,mitchell2019model,madaio2020co, metcalf2021algorithmic, gebru2021datasheets,rakova2023algorithms}. The research community's contributions around RAI Artifacts have so far focused on introducing new responsible AI artifacts~\cite{adkins2022method, hupont2024use}), examining how they are used within technology companies~\cite{deng2022exploring, balayn2023fairness,okolo2024you}), and interrogating the values embedded in their language~\cite{sadek2024guidelines}.






A subset of these artifacts, which we refer to as \textit{Multi-Actor Responsible AI (RAI) Artifacts}, intend to support different stakeholders, beyond those typically represented in an AI development team, in understanding AI-related information. In some cases, these artifacts intend to support \textit{direct} communication (e.g., dynamic conversation between or amongst different stakeholders), as in the case with compliance documentation such as \citeauthor{hupont2024use}'s Use Case Cards\footnote{Use Case Cards are designed for policymakers, researchers, and practitioners to assess risk level and document use cases~\cite{hupont2024use}}. Other times, they intend to support \textit{indirect} communication (e.g., communication of information to different stakeholders in isolation from each other), for example, in \citeauthor{mitchell2019model}'s Model Cards\footnote{Model Cards are promoted as an artifact that supports communication of model-related information to AI practitioners and audiences outside of the technology industry (e.g. journalists\url{https://modelcards.withgoogle.com/about}}. While a growing body of literature has emphasized the importance of multi-actor collaboration to support responsible AI design \cite{kaminski2018collaborative}, we have little insight into how well existing Multi-Actor RAI Artifacts foster the cross-sector communication essential for collaborative governance.  

In this paper, we center regulatory and civil society stakeholders to explore the perceived efficacy of Multi-Actor RAI Artifacts. Despite their central roles in governing the deployment of AI systems, their perspectives have so far been under-examined in studies on the design and use of Multi-Actor RAI Artifacts. As regulators begin to repurpose RAI artifacts as compliance tools, and civil society rely on transparent model documentation to examine industry practices, we anticipate that understanding these stakeholders’ perspectives 
will become increasingly important. We therefore ask: \textbf{RQ1.} \textit{How do legal/regulatory and civil society stakeholders perceive the efficacy of Multi-Actor Responsible AI Artifacts? } and \textbf{RQ2.} \textit{What are implications for improving the design and use of Multi-Actor Responsible AI Artifacts? 
}

To address these questions, we conduct semi-structured interviews and design activities with 19 participants involved in informing policymaking and advocacy around AI systems deployed by technology companies. During the activities, participants had an opportunity to directly examine existing filled-in examples of existing Multi-Actor RAI Artifacts to ground their reflections. The participants are from two focal stakeholder groups : (i) legal experts with prior or current engagement in the scholarship or practice of AI regulation; and (ii) individuals working for civil society organizations concerned with the impact of AI on society. 
Participants held beliefs about the role that RAI artifacts should play in society (i.e., protecting end-users from AI harms) that inherently conflicted with the goals they perceived the technology industry have for using RAI artifacts (i.e., promoting self-regulation and shaping societal standards for AI). Participants’ concerns surfaced how this underlying misalignment may ripple down to shape industry practices around the design and use of Multi-Actor RAI Artifacts. We organized participants’ concerns into four barriers: (1) end-users being positioned as second-order priority, (2) selective showcasing of laudable AI models versus truly high risk models, (3) over-reliance on transparency as a change mechanism, and (4) offloading duties for responsible use to end-users. Our findings describe how each of these barriers map to different motivating factors, artifact design and use practices, and downstream impacts on the broader society.

Overall, participants were concerned that, if left unaddressed, these barriers may (inadvertently) reconfigure power relations across civil society, government, and industry, impeding RAI governance efforts from those outside of the technology industry. Participants desired RAI artifacts that could, instead, foster more proactive and collaborative cross-sector AI governance efforts. We extend existing calls for broader community engagement in directing the role Multi-Actor RAI Artifacts play, particularly in supporting responsible AI efforts between and within the regulatory, civil society, and industry sectors. 

\section{Background}\label{Background}
The Responsible AI community has made fundamental strides in establishing core principles to guide the responsible development of AI. To instantiate these principles to practice, the community has developed new tools, processes, and resources. 
Throughout this paper, we use \textit{artifacts} as an umbrella term for these innovations. 
In this section, we review prior work on the evolving role of RAI artifacts in supporting multi-actor communication. We then review prior work on the design and use of RAI artifacts. 

\subsection{The Evolution of Artifacts as a Communication Stream for Industry, Regulation \& Civil Society}  
Artifacts have a longstanding history as vehicles to promote adherence to established values and principles critical to oversight, accountability, and overall governance---often in an inter-stakeholder capacity. In heavily regulated industries like healthcare, energy, manufacturing, aviation, and pharmaceuticals, artifacts such as checklists, risk assessments, and standard operating procedures facilitate communication and information flow among multiple actors (e.g. workers, managers, and external auditors) \cite{weske2019physician, weske2018using, ywakaim2019nurses, kagan2011fear}. These industries all have a shared history of adverse impacts affecting people, communities, and society-at-large.  The impacts have materialized as inequitable resource allocation \cite{blodgett2022responsible}, the perpetuation of systemic bias against social groups \cite{mehrabi2021survey}, and the unethical exploitation of vulnerable people to gain resources for the further enrichment of industry leaders \cite{noble2018algorithms}.

 In response, public pressure and advocacy across sectors have forced reflection within companies, pushed civil society members to drive major advocacy campaigns, and prompted a need for policymakers to introduce industry regulation to manage risk and error. Science historian Lorraine Daston posits that the identification, reduction, and elimination of errors has been a steady driving force of modern science influencing collective sensemaking and overtime shaping errors “as sites for collaborative, collective, and coordination work” \cite{lin2023bias}. To materialize this shared goal for risk reduction, artifacts have served as communication streams that encourage debate about power dynamics shaping the authority to define risks, influence public perceptions and normative standards for acceptable levels of risk, and forge new forms of partnerships by including multiple actors. In healthcare, checklists are used to indicate compliance with industry standards for patient safety and to produce documentation that allows third party auditing to affirm regulatory compliance needed for institutions to remain operational. Checklists are also used in architecture to avoid critical life-endangering errors \cite{bierska2022integrated}. In manufacturing, risk assessments are commonplace and often required by law promoting both short-term and long-term reflection on adverse impacts. The artifacts encourage communication among teams to discuss how to identify risks, they assist civil society advocates in holding companies accountable, and they communicate a level of compliance with enforced regulation.


Likewise, social computing scholarship has reflected many parallels. Researchers have extensively reported on AI risks, errors, and failures ~\cite{dominguez2023co, ntoutsi2020bias, dolata2022sociotechnical} and in response have created an abundance of RAI artifacts for assessment and mitigation, and to foster better communication between non-industry audiences who contend with the governance of AI. Unlike other industries, RAI artifacts have largely pre-dated AI regulation. Efforts to support the development of RAI artifacts have largely been centered on practitioners' needs.  Although RAI artifacts are commonly promoted for non-industry audiences, there is limited work incorporating the views of non-industry actors involved in AI governance. 
Some industry stakeholders have asserted that their technical expertise and domain knowledge uniquely positions them to foresee risks and build guardrails to mitigate those risks \cite{CHATTERJEE_BORDELON_2023, Levin_Downes_2023}. Meanwhile, proponents of regulation have been jostling to propose stricter government-led compliance and accountability to ensure industry stakeholders deliver on their promises of positive transformation and its ability to self-regulate \cite{ferretti2022institutionalist}. In a similar vein, Civil Society Organizations (CSO) have driven impactful change by leading grassroots efforts with impacted communities, grounding policy agendas in lived realities, and influencing AI practice \cite{stjernfelt2020role}. However, between total government regulation and total self-regulation, prior work suggests that the consensus will more than likely operate from a collaborative stance to take advantage of (rather than compete with) the varying expertise and abilities from the three sectors \cite{kaminski2018collaborative, Cohen_2023}.

\subsection{The Design and Use of Multi-Actor Artifacts in Responsible AI} 
In the past half decade, a growing community of responsible AI researchers and practitioners has advanced a range of RAI artifacts to help promote governance and increase transparency over development practices. For example, Accenture created the Fairness Tool to identify and mitigate bias; Google created Model Cards to support more transparent AI documentation; and Microsoft created Responsible AI Impact Assessments to help more proactively envision and mitigate downstream harms. This set of artifacts is expanding quickly, gaining widespread attention and use across AI application settings.  

The intended purposes and audiences of RAI artifacts differ. A subset of artifacts–like FairLearn~\cite{bird2020fairlearn} and InterpretML~\cite{nori2019interpretml}--are designed to support AI practitioners in quantitatively evaluating and improving AI systems along dimensions of common principles (e.g., fairness, explainability, accuracy). Another subset of artifacts—like Model Cards~\cite{mitchell2019model}, Datasheets for Datasets~\cite{gebru2021datasheets}, and Responsible AI Impact Assessments~\cite{metcalf2021algorithmic}–are also designed to support the improved design of AI systems, but focus on scaffolding \textit{documentation} of model information and \textit{reflection} on potential risks. For instance, Model Cards~\cite{mitchell2019model} supports AI teams in documenting model properties including training parameters and model evaluation results, and Datasheets for Datasets prompts documentation of dataset details, including the motivation, composition, intended uses of the dataset~\cite{gebru2021datasheets}. This latter class of artifacts serve broader functions of organizational, and increasingly, regulatory and societal governance of AI systems. In fact, based on documented information presented with these artifacts (e.g.,~\cite{mitchell2019model}), many of them are designed with an intention to be used by multiple different stakeholders, including auditors, policymakers, end-users, and other impacted AI actors who typically may not be part of the AI team developing the model. We refer to this specific class of RAI artifacts–those intended to support (direct or indirect) multi-actor communication–as \textit{Multi-Actor RAI Artifacts.} 

As Multi-Actor RAI Artifacts begin to play an increased role in supporting the governance of AI systems, their uses within the RAI ecosystem have also expanded and evolved. 
For instance, recently introduced regulatory regimes have pushed for the adoption of RAI artifacts by making completion of the artifact-supported process mandatory. The EU-AI Act and Algorithmic Accountability Act in the United States require that AI deployers complete an impact assessment prior to having a high-risk AI system enter the market~\cite{yoo2020regulation,novelli2023taking}. State legislatures in the United States have followed, proposing new AI bills that would require the developers and deployers of AI systems to complete an RAI artifact. The repurposing of RAI artifacts in these efforts are increasingly supported by RAI researchers, who have suggested that RAI artifacts are well-suited to support regulatory compliance~\cite{pistilli2023stronger}. 


With increasing multi-stakeholder interest and reliance on Multi-Actor RAI Artifacts, it becomes critical to understand how these cross-sector stakeholders perceive their efficacy. Existing work studying RAI artifacts have examined how they are used by industry AI practitioners~\cite{deng2022exploring, okolo2024you, balayn2023fairness}, what normative assumptions about ``AI ethics'' work they embed~\cite{wong2023seeing}, and the need to center \textit{effectiveness} evaluations--that is, an evaluation that examines causal relationships between the desired outcome of an artifact and their actual impacts in a given context~\cite{berman2024scoping}. However, minimal work has stepped back to consider what the ``desired outcome'' for RAI artifacts should be from the perspectives of cross-sector AI actors (e.g., regulatory and CSO actors), and whether and to what extent the existing design and use of artifacts help achieve these outcomes. Our study aims to address these gaps.

\section{Methods}\label{Methods}
\begin{figure*}
    \centering
    \includegraphics[width=0.90\linewidth]{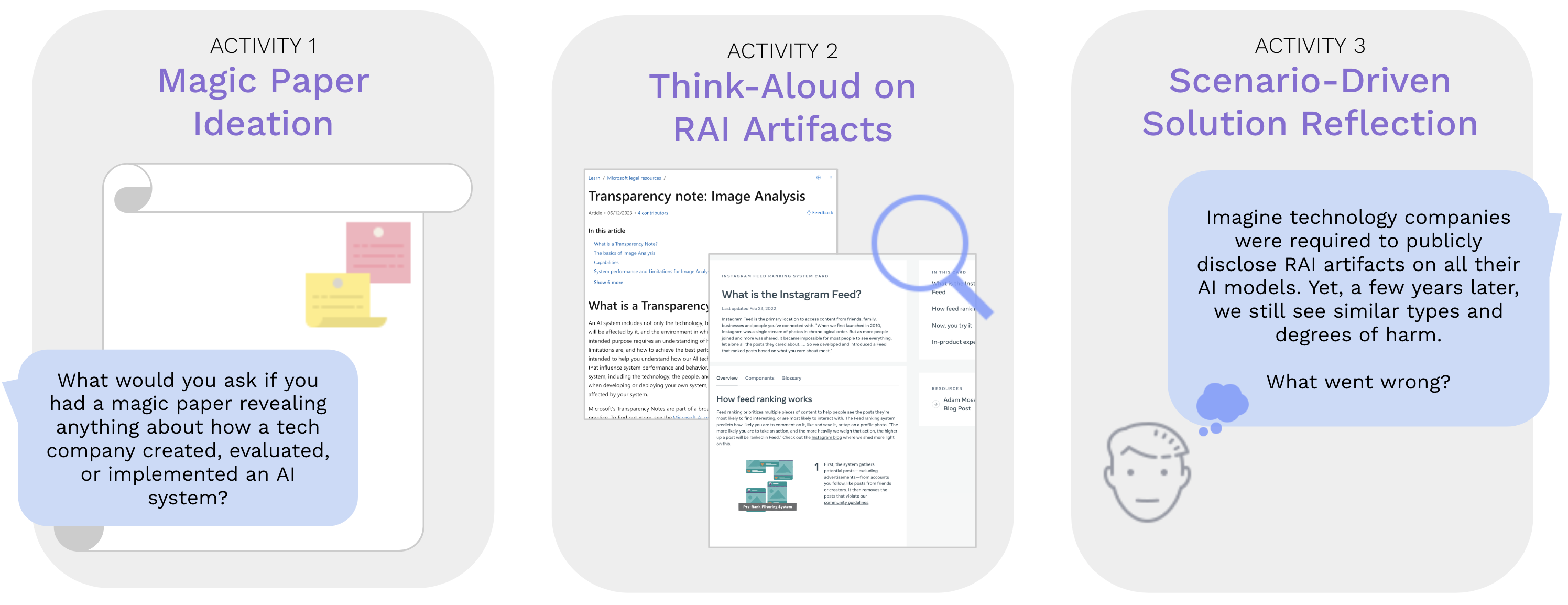}
    \caption{Overview of the three activities in our study protocol}
    \label{fig:activityoverview}
\end{figure*}

In this study, our goal was to understand how legal/regulatory and civil society organization (CSO) stakeholders perceive the efficacy of Multi-Actor Responsible AI Artifacts, including their desires for improving artifact design and use. Towards this goal, we conducted semi-structured interviews and design activities with 19 participants across these stakeholder groups.
Our study was approved by the institutional review board (IRB) at Microsoft. Below we discuss details about our steps we took to prepare for the study, our recruitment procedure, study protocol, and analysis approach.




\subsection{Study Preparation} \label{studyprep}
Prior to beginning our study, we aimed to contextualize our existing understanding of the broader landscape of RAI Artifacts, to inform the development of our study protocol. We therefore recruited and conducted formative interviews with four industry responsible AI practitioners who have experience informing the design of responsible AI artifacts.  The formative interview focused on (i) gathering input on the goals and intended uses of existing RAI artifacts and (ii) learning about industry practitioners' behind-the-scenes practices for ensuring their artifacts meet regulatory expectations and community needs. Knowledge shared during these formative interviews informed the creation of specific topics and questions in our study protocol, as well as our list of RAI artifacts to include in the design activity \footnote{The list includes: Data Cards (Google), Datasheets for Datasets (Microsoft), Model Cards (Google), Transparency Notes (Microsoft), System Cards (Meta).} The list includes a subset of publicly available RAI artifacts that were (i) created by technology companies and (ii) designed to be used by audiences from our stakeholder groups, based on publicly available documentation (e.g., research papers, company websites). We also completed pilot studies with three RAI researchers to help us identify additional points of improvement~\cite{majid2017piloting}. 

\subsection{Recruitment and Participant Background}
For our research study, we aimed to recruit individuals who had experience working in or with regulatory bodies and CSOs on topics related to governing the responsible development and use of AI systems. We first searched for government agencies and CSOs that are heavily involved in examining the use and impacts of deployed AI systems. After identifying a list of potential organizations, we looked on the organizations' websites for employed individuals working on relevant topics (e.g., technology policy or community advocacy). We also searched for individuals employed in academic institutions who had experience consulting for government agencies on topics related to AI regulation. We invited 34 potential participants to the study via email. While conducting each study session, we recruited additional individuals using snowball sampling. Overall, 23 stakeholders responded to our recruitment emails and 19 participants were scheduled for interviews. 

All 19 participants indicated having prior experiences informing regulation and public policy around AI systems in government or civil society, as well as having experiences investigating the downstream societal or legal impacts of industry-created AI systems. 
We report on participants' stakeholder grouping based on their current organizational affiliation. Our sample represented stakeholders across thirteen organizations of varying sizes, residing in both CSOs and regulatory bodies, representing four global locations (U.S., E.U., Mexico, and Brazil) with 47\% identifying as men, 36\% identifying as women, and 17\% preferring not to disclose. We include participant demographics and details of their background in Table \ref{table:demo}.    

After conducting studies with 19 participants, no new themes emerged, indicating that a point of saturation was achieved \cite{saunders2018saturation}. The remuneration for the study was \$50 USD  (or equivalent amount in the participant's local currency) which was distributed via an online gift card. All interviews were conducted online using a video conferencing platform between July and October 2023. 
\begin{table*}[]
\centering
\scalebox{0.8}{
\begin{tabular}{lllll}
\toprule
Participant ID & Professional Role &Stakeholder Group                    & Years of Experience      & Location \\
\midrule
P01            & Researcher & Civil Society                               & 1-   3 Years             & U.S.     \\
P02            & Attorney & Law/Regulation                                 & \textgreater 7 Years     & U.S.     \\
P03            & Attorney & Law/Regulation                                 & \textless 1 Year         & U.S.     \\
P04            & Researcher  & Civil Society                             & 1- 3 Years               & E.U.     \\
P05            & Researcher  & Civil Society                              & 1- 3 Years               & U.S.     \\
P06            & Attorney; Community Advocate         & Civil Society      & \textless 1 Year         & U.S.     \\
P07            & Researcher & Civil Society                               & 3 - 5 Years              & U.S.     \\
P08            & Attorney; Community Advocate   & Civil Society            & \textgreater 7 years     & U.S.     \\
P09            & Researcher; Community Advocate  & Civil Society          & 5 - 7 Years              & U.S.     \\
P10            & Researcher   & Civil Society                             & 1- 3 Years               & U.S.     \\
P11            & Community Advocate & Civil Society                       & 1- 3 Years               & U.S.     \\
P12            & Civil Society / Civil Rights Technologist & Civil Society & 1- 3 Years               & U.S.     \\
P13            & Researcher; Community Advocate  & Civil Society           & 5 - 7 Years              & U.S.     \\
P14            & Researcher; Analyst & Civil Society                       & \textless 1 Year         & U.S.     \\
P15            & Attorney  & 
Law/Regulation                                & 3 - 5 Years              & U.S.     \\
P16            & Researcher  & Civil Society       & \textgreater 7 Years     & Mexico   \\
P17            & Attorney; Researcher & Law/Regulation                      & 5 - 7 Years              & U.S.     \\
P18            & Attorney; Researcher & Law/Regulation                        & \textgreater 7 Years     & U.S.     \\
P19            & Attorney   & Law/Regulation                                & \textgreater 5 - 7 years & Brazil \\
    \bottomrule
\end{tabular}
}
\caption{Demographics of study participants} \label{table:demo}
\end{table*}
\subsection{Study Overview}
We conducted 60 minute study sessions with each participant, including a semi-structured interview and design activity. We began each study session by asking the participant to contextualize their question responses within their own past experiences, role-assigned responsibilities, and goals around responsible AI. By tying their responses to their background as a regulatory or civil society stakeholder, we aimed to better understand how their goals, beliefs, and experiences shaped their perceptions of RAI artifacts. After asking related questions about their RAI goals and work, we went through three core activities: (1) Magic Paper Ideation, (2) Think-Aloud on RAI Artifacts, and (3) Scenario-Driven Solution Reflection. Figure~\ref{fig:activityoverview} includes an overview of the three activities and the Appendix~\ref{appendix:interviewguide} includes the question guide. 

\subsubsection{Activity 1: Magic Paper Ideation} \label{magicpaper}
Our goal for the Magic Paper Ideation Activity was to learn about participants' information needs and desires for RAI artifacts. In particular, we elicited each participant's ideas about the types of model-related questions they wished an ``ideal’’ RAI artifact would address. Figure~\ref{fig:activityoverview} includes an exemplar question. 
The participant's ideation work completed during this activity helped prepare them for the next activity, in which they were asked to examine an example of an existing RAI artifact. 

\subsubsection{Activity 2: Think-Aloud on RAI Artifacts} \label{thinkaloud}
This activity helped ground participant reflections and assessments of artifacts using actual examples. At the start of this activity, each participant was shown a list of Multi-Actor RAI Artifacts (see Study Preparation). 
After reviewing the artifact list, the participant was asked to select one to delve deeper into. 
We then provided the participant a link to a filled-in example of the artifact they had selected. We asked the participant to think aloud their thoughts and reactions to the artifact, as they inspected it in their own browser. To better understand \textit{what} specific components of the RAI artifact led to their reaction, \textit{why}, and improvements they desired, we asked follow-up questions as needed. 
Figure~\ref{fig:activityoverview} includes two examples of filled in-artifacts.

\subsubsection{Activity 3: Scenario-Driven Solution Reflection} \label{reflection}
In the Scenario-Driven Solution Ideation activity, we presented hypothetical scenarios to help prompt reflection around alternative artifact designs and processes that participants believed could help technology companies be accountable for their AI systems. We first asked participants to reflect on our prior discussions about the limitations they believed existing artifacts had. We then presented the scenarios to help scaffold ideation. Figure~\ref{fig:activityoverview} includes an example scenario.


\subsection{Data Collection and Analysis}
After asking participants for their consent, we audio recorded and took notes from the design activity. All 19 participants consented to both recording audio and note-taking. Audio recordings were transcribed automatically and manually checked for accuracy. To qualitatively code the transcriptions, we followed a reflexive thematic analysis approach~\cite{braun2019reflecting}. To assist in the coding process, we used HeyMarvin\footnote{HeyMarvin (\url{https://heymarvin.com/}) is an online platform that offers analytical tools for qualitative data with features to code collaboratively as a research team.} as a tool to streamline collaborative coding among the three authors. The overall coding process included analyzing data through multiple rounds of coding and iterative affinity diagramming. 

To code the transcripts, the first author familiarized themselves with the data and identified initial lower level codes for a subset of the transcripts. The first author then discussed the codes with the second and third authors to discuss disagreements and calibrate on the coding style and granularity. 
Any disagreements were resolved through synchronous discussions. The lower level codes were then transferred to a shared online board for collaborative affinity diagramming~\cite{beyer1999contextual} using Figma, a web application where multiple users can synchronously generate and arrange sticky notes. The three authors iteratively grouped the first-level codes into successively higher level codes (i.e., second and third level codes). After completing each additional level of coding, the authors collaboratively discussed emerging themes and relationships between codes~\cite{braun2006using, lazar2017chapter}. 

\subsection{Limitations}
We acknowledge the methodological limitations of our study. Our sample was largely skewed toward US-based participants. To diversify the participant sample, we recruited participants broadly across stakeholder groups, demographics, and location. Throughout the recruitment process, we monitored the participant demographics represented in our participant sample to guide decisions about which populations to refocus recruitment efforts towards. 
Second, it is possible that, because we are employees within a technology company, our participant sample is skewed towards those who are supportive and optimistic of industry responsible AI efforts. We have reason to believe this because we experienced instances where recruitment contacts objected to study participation due to moral discomfort with engaging in research funded by the technology company. We respect their right to express that stance, and acknowledge their freedom to choose to voluntarily participate. We encourage future research that extends this line of inquiry in different sectorial and geographic contexts, where expectations around responsible AI goals and needs may differ. 


\section{Results}\label{Results}
\begin{figure*}
    \centering
    \includegraphics[width=0.95\linewidth]{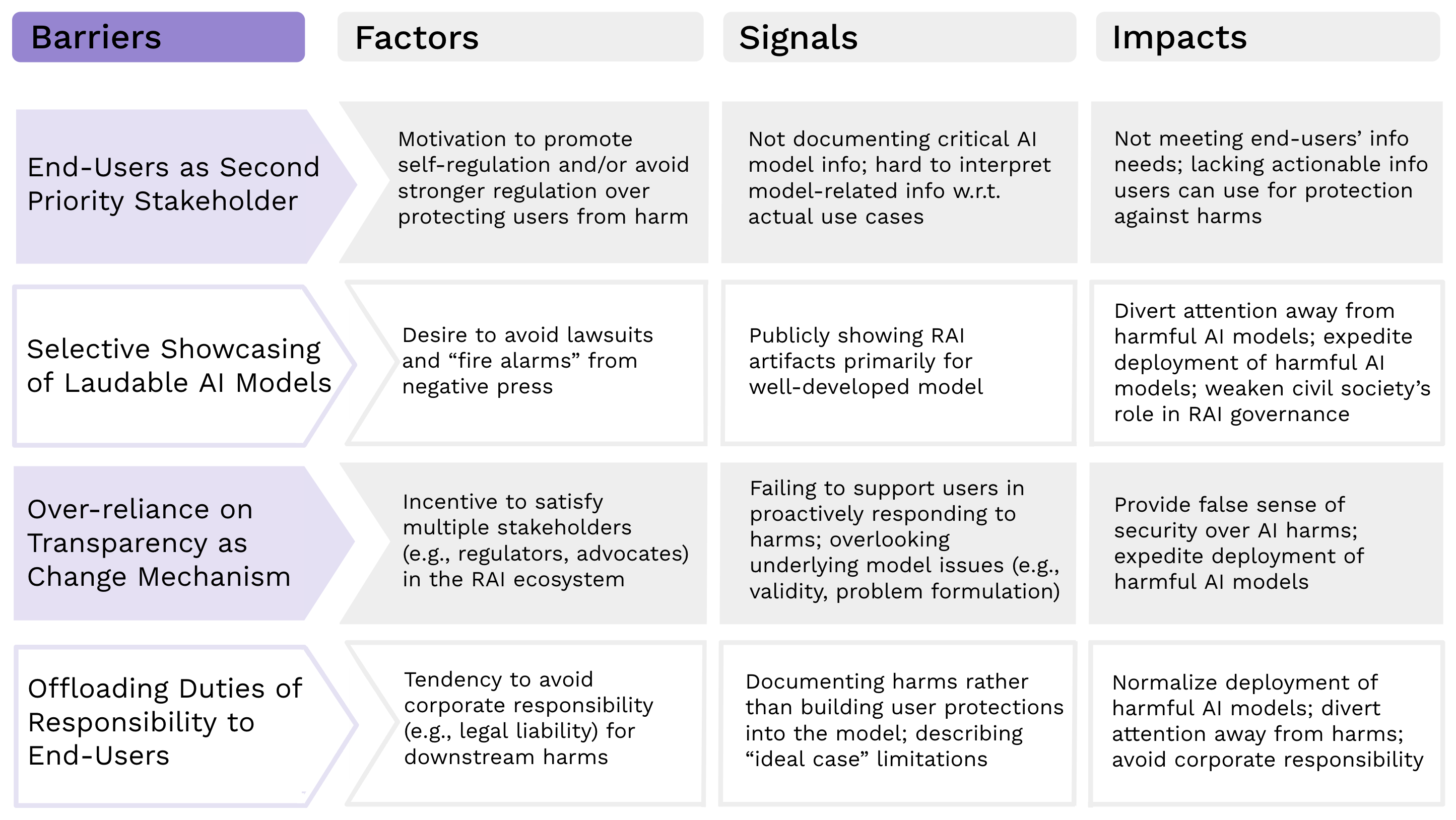}
    \caption{A summary of the four barriers to Multi-Actor RAI Artifacts surfaced by participants, each mapping to one subsection of the findings. For each barrier, we describe (1) Factors: the incentives, goals, and theories of change of the technology industry that contribute to the barrier, (2) Signals: ways in which the barriers become visible through design and use practices around RAI artifacts, and (3) Impacts: the broader downstream impacts the barriers have on the responsible AI ecosystem and society.}
    \label{fig:barriers_table}
\end{figure*}

While participants recognized the importance of artifacts in the RAI ecosystem, our findings focus on communicating their concerns around the potential role and impacts of RAI artifacts. As described in Figure~\ref{fig:barriers_table}, participants’ reflections on and assessments of existing Multi-Actor RAI Artifacts surfaced four barriers they believed impeded their ability to protect end-users from downstream AI harms: (1) end-users\footnote{Here and throughout the paper, ``end-users'' refers to stakeholders who may be impacted by the downstream use of an AI system, including those who may use the AI model or those who may be impacted by decisions made with the AI model.} being positioned as a second-order priority in the design and use of RAI artifacts, (2) intentionally showcasing the RAI artifacts for a select set of laudable AI models, (3) over-relying on transparency as a change mechanism when designing RAI artifacts, and (4) designing RAI artifacts to delegate duties for responsible use to end-users. We organize the findings subsections around these four barriers, with each subsection describing participants' perspectives about the technology industry's goals and incentives that contribute to the creation of the barrier, the design and use practices of RAI artifacts that make the barrier visible, the downstream impacts that barrier has on the broader RAI ecosystem, and corresponding desires participants had for improving the role, design, and use of RAI artifacts. 


\begin{figure}
    \centering
    \includegraphics[width=0.9\linewidth]{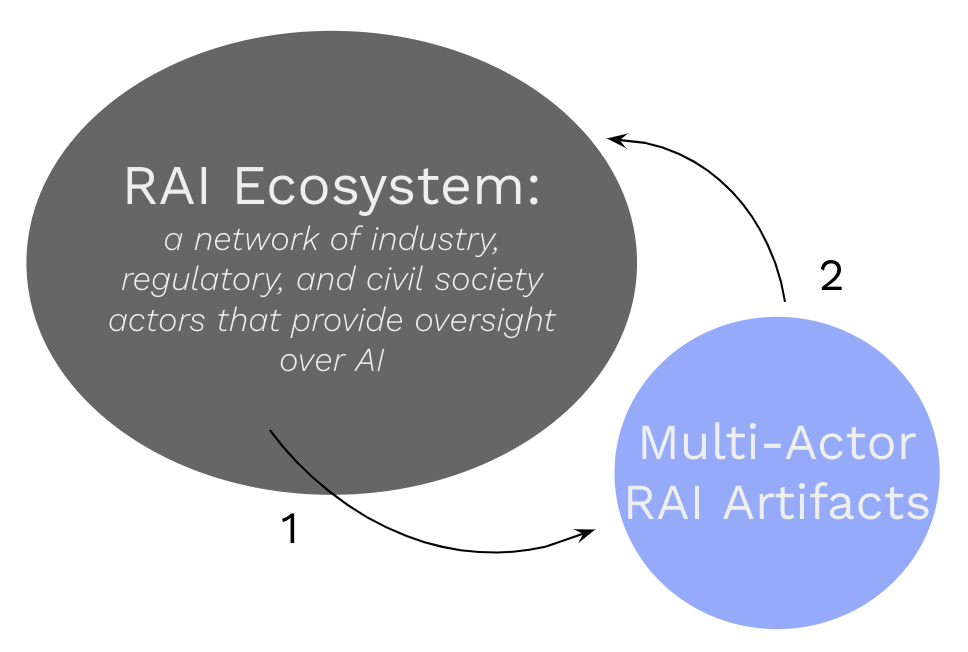}
    \caption{Participants described how (1) the technology industry exists within a multi-actor RAI ecosystem, which shape their goals and incentives for RAI governance. These goals and incentives (2) shape how the technology industry designs and uses Multi-Actor RAI Artifact; these decisions have rippling impacts on other actors within the RAI ecosystem. A more detailed breakdown of the properties and impacts of this dynamic are described in Figure 3.}
    \label{fig:ecosystem}
\end{figure}
\subsection{End-Users as Second Priority Stakeholder} \label{endusers}
A key theme that emerged revolved around how Multi-Actor RAI Artifacts (perhaps unintentionally) position end-users as secondary stakeholders.  This can occur, for instance, when information documented in RAI artifacts is shaped in ways that are ultimately \textit{inaccessible} and \textit{inactionable} to impacted end-users. While RAI artifacts are currently marketed as being created for a wide range of stakeholders\footnote{For example, Model Cards is marketed towards broad audiences (both ``experts and non-experts'') including developers, journalists, and industry analysts~\url{(https://modelcards.withgoogle.com/about)}. Datasheets for Datasets is marketed towards a wide range of actors who contend with datasets including practitioners and academics.}, including impacted end-users and non-technical stakeholders, participants observed that the \textit{type or presentation of content} is tailored towards technology companies' needs rather than the needs of impacted community members. Participants acknowledged the challenges with designing \textit{``one size fit all''} RAI artifacts and, instead, advocated for greater effort invested in designing RAI artifacts that center impacted end-users' domain-specific informational needs. 

For example, participants perceived a misalignment between the type of content presented in RAI artifacts versus the type of content that would be valuable for protecting community members from AI harms. 
In some cases, RAI artifacts appeared to omit details that would be critical for impacted community members. As one participant articulated, RAI artifacts like Model Cards that center accuracy-based performance measures overlooks other harms that may arise from the downstream use of the AI model: 

\begin{quote}
``I'm so glad that they are addressing some of the harms that were articulated in Gender Shades. However, I think for a lot of community-based organizations that I've talked to, it would be a frightening outcome if the harms that they're worried about or simply considered [only includes] accuracy[\dots ] A more accurate Big Brother isn't safer or better.'' (P05, Researcher, Civil Society) 
\end{quote}

\vspace{-0.5mm}
Participants also observed that many RAI artifacts actively discussed in industry RAI efforts provide upstream information about the AI model (e.g., dataset properties, performance on training data). However, they believed that civil society and impacted communities would find information contextualized to the AI model's downstream use cases more meaningful. As one participant articulated, the current design of some RAI artifacts ``divorces the conversation from the uses of the model, which is where the harms happen'' (P02, reflecting on Model Card on Face Recognition Algorithm). Examining Facebook's System Cards for Instagram Feed algorithm, another participant noticed that the artifact only showed a subset of their rules for how the algorithm selects posts to show. The participant wondered whether the company strategically decided to provide examples of rules that are the least problematic: ``I love that they gave an example, but\dots I want to know what are all the rules \dots usually they select a very mundane [rule] that's super nice and, you know, [one that is] simple and wouldn't raise any red flags'' (P10, Researcher, Civil Society reflecting on System Cards for the Instagram Feed Algorithm). The participant further elaborated that they desired complete information on the AI model to more critically examine the potential societal implications of the algorithm's design decisions: ``I want to see the full list of rules, so I can say, `OK, these rules are great. This rule might cause these harms for these and these groups.''' (P10,  Researcher, Civil Society, reflecting on System Cards for Instagram Feed Algorithm). Participants were concerned because this information omission prevented end-users from accessing details that could empower them to better reason about or protect themselves from potential harms. 

Moreover, several participants emphasized that this barrier was an inherent byproduct of technology companies’ motivations to leverage RAI artifacts as a tool to shape attitudes and industry norms. As one participant who informs technology policy and regulation at a CSO described, Multi-Actor RAI Artifacts are used tactically to promote self-regulatory practices and avoid stricter AI-related regulation: ``[RAI artifacts] help head off stronger regulation. You know companies saying, `oh hey, we already have these standards. Just look at these standards.'\dots I think it is very much driven by a desire to head off stronger regulations, perhaps’’ (P08, Attorney, Civil Society). As a result, even if RAI artifacts are designed with an intention to support model understanding for multiple actors, including end-users, there are disincentives embedded within industry norms that prevents prioritizing user needs in the design of RAI artifacts. 
\subsection{Selective Showcasing of Laudable AI Models}
\label{results:selective}


Participants were concerned that technology companies strategically showcased RAI artifacts for well-developed, often lower risk AI models--in part, as a tactic to improve the company's public image--while avoiding showing artifacts for models that would benefit the most from external scrutiny and feedback. Participants emphasized the importance of showing external stakeholders RAI artifacts for potentially controversial or ``work in progress’’ AI models, so they can use them to identify and surface problems that may arise. Participants pointed out that this created a paradox in efforts to develop Multi-actor RAI Artifacts: Even though Multi-actor RAI Artifacts are marketed as a mechanism to promote cross-stakeholder communication, technology companies are least likely to disclose RAI artifacts for the AI Models that would benefit the most from this form of external information sharing. As one participant articulated: 
\begin{quote}
``I could also see a company releasing artifacts[\dots ] where it makes them look good and they don't see problems. But then the ones that are problematic are not going to be put forth publicly, and you might distract civil society and researchers by focusing on the things that you have put out there[\dots ] sort of direct people's energies away from the ones that are a work in progress.'' (P02, Attorney, Regulation)
\end{quote}

Participants acknowledged that companies may be hesitant to disclose RAI artifacts for all their models, in large part, because there are currently no safe mechanisms for companies to do so. Participants reflected on how conflicting theories of change between civil society and the technology industry pose natural tensions to realizing disclosure policies favorable to civil society. For example, some participants described how, in some countries (e.g., United States), civil society relies on raising ``fire alarms'' by publicly shaming or otherwise pressuring technology companies into improving their AI development practices. As one participant reflected, RAI artifacts failing to prevent downstream harms from AI systems (by failing to showcase work-in-progress AI models) may be a ``feature of the [US] system'' (P08). Others pointed out that legal teams at technology companies are unlikely to allow disclosure of artifacts that surface potential model harms. As one participant acknowledged: ``companies are really disincentivized from self disclosing these kinds of risks because their own legal departments are worried about liability if they claim to be aware of a certain kind of harm.'' (P05, Researcher, Civil Society) 

Reflecting on the information types included in Model Cards, another participant further elaborated that these liability challenges may disincentivize companies from documenting information about AI models that could help end-users protect themselves from potential harms: 
\begin{quote}
``[harm use cases] is just not information that they want to connect to their technology [\dots ] because [even if] they don't intend this use [\dots  if] people are using it in this way [\dots ] it both shows acknowledgement of harmful uses and maybe an acknowledgement of unintended uses'' (P02, Model Cards on Face Detection Algorithm) 
\end{quote}


Because of these dynamics, participants were concerned that technology companies may divert public attention from the limitations and risks of their AI systems by ommitting information that may promote conversations about AI harms. Other participants were concerned that this may, in turn, weaken the role that civil society can play in identifying and mitigating these harms. Many civil society actors therefore acknowledged the importance of utilizing artifacts to study how downstream harms affect communities but they were eager to \textit{``study up''} \cite{nader1972up} by learning more information about upstream power dynamics, structures and decision-making that contribute to more systemic challenges in AI. In particular, participants wanted Multi-Actor RAI Artifacts to document the \textit{social and organizational context} surrounding technical decisions and \textit{value-based judgements, assumptions, and policy decisions} made during AI development. For instance, participants pointed out that information around \textit{who} was involved in model development (from problem formulation to evaluation), how organizational incentives may have impacted AI development decisions (e.g., metrics used for promotion), or even how much data labelers were paid could help surface organizations' underlying values and priorities around AI development. 

Other participants pointed out that, while the RAI Artifact they examined documented technical details about the AI model, they were more interested in ``what policy choices are embedded in seemingly technical decisions'' (P02). Participants described the value of documenting decisions from earlier on in the AI development process, including who framed the problem space that the AI model intends to address. As one participant elaborated: 

\begin{quote}
    ``Just knowing what the purveyors of the system or the developers of the system thought that they were solving for is illuminating. Because again, sometimes it might just be, `Well, there wasn't really a problem, but we have the capacity to do this. It's efficient. We think of tech as better always and so we just built it.''' (P02, Attorney, Regulation) 
\end{quote}

\vspace{-1.5mm}
Some participants also believed that, in disclosing RAI artifacts for work-in-progress models, artifacts can help promote more proactive collaboration between civil society and the technology industry. Participants recalled that they sometimes provide consulting services to technology companies to share early-stage feedback for their AI models, but that these efforts also suffer from the same paradox: Companies are more likely to consult civil society on AI models that they know are already well-developed and may be disincentivized from engaging them in conversations around AI models relying on disagreeable problem formulations. Therefore, participants advocated for new processes and policies that can provide safe pathways for technology companies to solicit feedback from civil society earlier on in their development process.

\subsection{Over-Reliance on Transparency as a Change Mechanism}
\label{results:transparency}
In reflecting on why participants found it critical for civil society stakeholders to use RAI artifacts to ``raise alarm bells’’ that prompt real-world change, participants described that transparency in itself typically does not help mitigate or address harm. However, they also acknowledged that proposals for greater transparency were a safe ``political’’ first step that could help reach a middle ground across civil society, government, and industry actors. As one participant reflected: 

\begin{quote}
``[Data transparency] is, in some sense, the least controversial thing that policymakers [and] civil society can ask for, that will piss the least people off, and that, frankly, [is] the least way to get companies to actually do anything right [\dots ] You end up pissing the least people off politically when you are calling for transparency because it's something that everyone can agree upon.'' (P07, Researcher, Civil Society) 
\end{quote}

Even though it may be more challenging to push forth governance mechanisms ``with teeth'', many participants emphasized the importance of exploring these next steps. As the participant further elaborated: ``it's not always clear what, then, gets done with that transparency. There's a lot of, `sunlight will fix the problems.''' (P07, Researcher, Civil Society). Instead, one participant described that simply having potential harms documented may normalize their existence: ``The interesting thing [\dots ] I tend to feel about transparency [\dots ] looking at the information that I have in front of me, and like not caring about it because it's there, you know?'' (P07) 

Similarly, participants were particularly concerned about the continued of RAI Artifacts to help placate the image of AI models and products that they viewed as being fundamentally problematic. For example, examining the Transparency Notes for a sentiment analysis model, one participant described: 

\begin{quote}
``It's such a problematic product and you know it. It's like, `oh, here are some key limitations to consider.' And those limitations are so major and yet the product is being sold. You know, so I don't think if you're selling a product that can be inherently problematic [\dots ] a statement of some limitations is really sufficient.'' (P15, Transparency Notes for Sentiment Analysis feature) 
\end{quote}


Highlighting patterns of inappropriate use and unanticipated impact, some participants believed that RAI artifacts may, in fact, negatively contribute to concerning trends in the types of AI systems that are deployed. In particular, participants conceptualized how RAI artifacts may serve as buoys for technology companies, allowing them to keep harmful AI systems afloat without addressing their harms. As one participant hypothesized, these dynamics may also expedite the deployment of new AI systems: 

\begin{quote}
    ``Really, really powerful technology is going out the door with very few guardrails except for this like quite thin amount of documentation [\dots ] think it's a pretense that like a lot of systems have been expedited out the door with this in place. [It gets] the boxes checked.'' (P05, Researcher, Civil Society reflecting on Model Cards for Face Detection Algorithm) 
\end{quote}

As such, participants described that they desired RAI artifacts that empower external stakeholders to not just relay, but \textit{respond} to and help mitigate potential harms. 

\subsection{Offloading Duties of Responsibility to End-Users}
\label{results:offloading}
Examining Multi-Actor RAI Artifacts that center the disclosure of harms, participants further expressed that technology companies may be leveraging artifacts to set expectations for responsible use and offload the burden for responsibility on to end-users. Participants described ways in which the content design of the RAI Artifact appeared to protect the company from liability risks, at the expense of protecting users from potential harms. 

For example, participants expressed concern that, even in Transparency Notes, which is designed exclusively for impacted end-users, the information shared about the AI model negatively shapes expectations over \textit{who} is responsible for potential downstream harms. Reading the starting paragraph of a Transparency Notes document\footnote{``Notes are intended to help you understand how our AI technology works, the choices system owners can make that influence system performance and behavior, and the importance of thinking about the whole system, including the technology, the people, and the environment.''}, one participant reacted: 

\begin{quote}
``I think the note in the second paragraph, right away, raises a little bit of a red flag for me [\dots ] Basically like [it's saying, `if] you're going to use our system, [it] is your responsibility to make smart choices.' As opposed to saying, `here are the choices we made that will then impact what you can or cannot do with this system.'' (P13, Researcher/Advocate, Civil Society) 
\end{quote}

Similarly, another participant was surprised by the title ``Transparency Notes'', expressing: ``my initial response to Transparency Note [was] `Oh, Microsoft is being transparent about something.' But what it feels like they're doing here is putting the responsibility on someone else who is going to be using the technology'' (P10, Researcher, Civil Society). 

Another participant examining Transparency Notes pointed out that the example the artifact used to convey model limitations\footnote{The participant was referring to the following text on the Transparency Note: ``For example, that was awesome, could be either positive or negative depending on the context, tone of voice, facial expression, author and the audience.''} appeared to communicate limitations in an ideal case scenario, rather than the limitations that end-users may experience in practice. The participant elaborated that it would instead be helpful to provide examples that more closely mirrored the real-world limitations in the model that users may encounter: 

\begin{quote}
    ``If I say `you're such a boss a** b**ch', that's a good thing, right? But if I say you're such a b**ch, that's a bad thing. And that could easily be an example that's included here. It doesn't sound very professional, but it really illustrate extent to which it's so sensitive [\dots ] it just makes it clear that this [limitation is] kind of idealistic, and it won't really be able to measure things.'' (P16, Researcher, Civil Society, reflecting on Transparency Notes on Sentiment Analysis feature)
\end{quote} 

As P16 described, and as Section~\ref{endusers} elaborates, participants generally desired RAI artifacts that prompted technology companies to disclose realistic model-related information that would could be more easily interpreted by end-users, even if this means that the model creator (e.g., the technology company or AI team) may not look as reputable. 

\section{Discussion}\label{Discussion}
Participants’ reflections on existing RAI artifacts surface how Multi-Actor RAI Artifacts reside within a broader AI governance ecosystem, shaping their design and use. Figure~\ref{fig:ecosystem} shows how the technology industry’s goals, incentives, and theories of change shape how they design and use RAI artifacts. Participants perceived their own RAI goals (i.e., protecting end-users from AI harm) as conflicting with that of industry (i.e., promoting self-regulation and shaping societal standards for AI), leading to concerns about the broader downstream impacts of these practices. 

We organize participants’ concerns into four key barriers that help break down participants’ beliefs about the current overlooked impacts of Multi-Actor RAI Artifacts: (1) End-users as second-order priority, (2) Selective showcasing of laudable AI models, (3) Over-reliance on transparency as a change mechanism, and (4) Offloading duties for responsible use to end-users. To overcome these barriers, participants desired structural changes to the RAI governance ecosystem, coupled with improvements in how RAI artifacts are designed and used. Drawing on these findings, we discuss implications for research, design, and policy.

\subsection{Challenges from Conflicting Theories of Change}
Analogues to the kinds of processes and artifacts we study in the AI context have existed in mature industries like transportation, agriculture and pharmaceuticals for decades.  Indeed many RAI artifacts take direct inspiration from consumer- and practitioner-facing documentation such as nutrition labels and checklists.  However, whereas in the more mature industries artifacts generally arose \textit{in response to} existing regulation, and were de facto instruments for compliance \cite{boon2010front}, RAI artifacts largely predate regulatory guidance on AI.  As our participants noted, RAI artifacts play a major role in \textit{shaping} expectations and standards around AI in the first place.  


Within these dynamics, we can observe how actors across sectors (e.g., regulatory, civil society, industry) each have their own—often implicit—desired outcomes, incentives, and strategies they use to advance their vision for what responsibly governing AI should or could look like. In our study, participants drew on these conflicting theories of change to both explain their own assessments of the efficacy of artifacts, as well as explain why technology companies currently leverage artifacts. In particular, participants described how their judgements about the technology industry’s underlying goals for RAI mediated their perceptions of the efficacy of Multi-Actor RAI Artifacts. Participants believed that their own desired end-goals for RAI artifacts (i.e., protecting end-users from AI harms) deviated from technology companies’ underlying end-goals (i.e., promoting self-regulatory practices and shaping societal expectations around AI). They identified how these clashing goals lead to the manifestation of four barriers that impede the effective design and use of RAI artifacts, where ``efficacy'' is judged with respect to their own end-goals. 

At first glance, these tensions pose significant challenges. Many of the challenges we discussed are a natural consequence of the technology industry operating off of a self-governing model under a capitalist value system, indicating they may be nearly impossible to mitigate without substantial societal changes. For example, technology companies will tend to err on the side of not disclosing harmful model-related information, for concern of backlash and profit-related harms, even if this may be counter-productive to the goal of proactively identifying and mitigating AI harms to impacted stakeholders. However, we argue that these barriers are not \textit{inevitable} in broader societal efforts to govern AI. Future work should investigate structural changes that could be made through new regulation and advocacy efforts, that could support the need for \textit{meaningful disclosure} amongst regulatory and CSO stakeholders. We continue this discussion in the next section on design and policy implications. 

\subsection{Towards Facilitating Multi-Actor Communication}


In our findings, we observe barriers to the responsible design and use of RAI artifacts that hinder multi-actor efforts around AI governance. These barriers ultimately reduce legal, CSO, and impacted end-users' ability to proactively anticipate or mitigate AI-related harms. Borrowing from the social sciences, this can be described as reducing \textit{social capacities}--the ``abilities, skills and competences that help people to better prepare for, respond to, recover from or adapt to the negative impacts of [\dots] hazard events''~\cite{hoppner2012linking, kuhlicke2010knowledge}. \citeauthor{hoppner2012linking} argued that improved risk communication is vital for crafting environments that help build up communities' social capacities \cite{hoppner2012linking}. In practice, risk communication has been a hallmark feature embedded in design and policy around Multi-Actor RAI Artifacts; it is also increasingly a fiduciary responsibility for high-risk use cases of AI with the emergence of newer regulatory frameworks~\cite{calvi2023enhancing}. However, as our findings suggest, industry-external stakeholders have been overlooked in efforts to communicate AI-related risks from RAI artifacts. This paper emphasizes the importance of questioning \textit{for whom} RAI artifacts are designed, to in turn deliberate \textit{which risks} and \textit{how} these risks are (not) communicated. 

Towards centering these considerations, we extend existing calls to increase stakeholder participation (particularly those residing outside of the technology industry) in the design of RAI artifacts. 
As our findings describe, treating these stakeholders as ``secondary’’ stakeholders risks overlooking their unique needs and desired uses of RAI artifacts.  
For example, participants described how use cases and limitations for AI models should be grounded in what they experience in their day-to-day, to help them anticipate certain harms. 
Moreover, these information needs may differ depending on the type of AI model covered in the artifact. 
Future efforts that use RAI artifacts for governance should account for the type of AI model, like the associated risk level (e.g., as in the EU AI Act~\cite{madiega2023artificial}) or level of interaction with end-users, to assess whether and when certain RAI artifacts can support governance. This may require scoping down the intended uses of a given RAI artifact, countering the industrial tendency to design for scalability~\cite{hanna2020against}.  
Besides designing RAI artifacts to support improved communication of AI model information to stakeholders, participants wanted artifacts to better facilitate \textit{collaboration} amongst industry, regulatory, and CSO stakeholders. For example, CSO participants described that accessing RAI artifacts for ``work-in-progress'' AI models could allow them to play a bigger role in risk identification and mitigation, before models are deployed. Future work could explore how policy interventions can incentivize technology companies to regularly involve CSO stakeholders as third party auditors of their AI systems, especially in earlier stages of AI design.

\section{Conclusion} \label{Conclusion}
Multi-Actor Responsible AI Artifacts can play a pivotal role in fostering collaborative AI governance efforts amongst industry, legal, and civil society actors. This study advances our understanding of how to realize this vision by addressing an outstanding gap in literature: Understanding legal and civil society stakeholders' perspectives around the efficacy of existing Multi-Actor RAI Artifacts.  

\section{Acknowledgements}
This research was conducted at Microsoft Reseach. We thank our participants for sharing their valuable time, insights, and experiences for the study, and the reviewers for sharing constructive feedback. We are also very grateful for the feedback shared by Ryan Steed, Seyun Kim, Alicia DeVrio, Mihaela Vorvoreanu, and those in the Microsoft Research FATE group and Office of Responsible AI. 

\section{Reflection on Ethics and Social Impact}

\subsection{Positionality and Adverse Impact Statement}

In this work, our positionality stems from a multidisciplinary perspective at the intersection of technology, ethics, and society. As researchers, we acknowledge our own subjectivities, backgrounds, and positionalities, which inevitably shape our interpretations, methodologies, and contributions to the study.

Our team comprises individuals with diverse academic and professional backgrounds, including computer science, human computer interaction (HCI), and responsible AI. This diversity enriches our research by providing varied perspectives and insights into the complex issues under consideration in our study. While not paid explicitly to complete this project, the authors were supported by a large technology company in the United States. We recognize the inherent power dynamics embedded within the research process and opted to mitigate biases and adopt inclusive guidelines throughout our methodology to ensure participants were recruited respectfully while being mindful of potential privileges.

Our goal in conducting this research was to incorporate legal and civil perspectives around the use and design of Multi-Actor RAI Artifacts. However, in adopting this framing it is possible that criticisms surrounding existing artifacts could be misconstrued as advocating for their removal or limited use. Instead, we emphasize the important and fundamental role of current Multi-Actor RAI Artifacts and envisions utilizing the insights presented to charter a wave of improvements to facilitate improved communication among RAI actors.  

In presenting our findings, we acknowledge the limitations of our positionality and strive to be transparent about our methodologies, assumptions, and potential biases. We invite readers to critically engage with our work, interrogate our perspectives, and contribute to ongoing conversations about the role of Multi-Actor RAI Artifacts in the field.

\subsection{Participant Safety and Ethical Considerations} 
Throughout the study, we reminded stakeholders that their participation was completely voluntary and that their responses would be kept anonymous. To minimize risks of identification, we report on demographics from an aggregate level, assign unique identifiers in our reporting, and opt to report on stakeholder groupings opposed to reporting on participant affiliations. For the study sessions, participants were given the agency to consent to their audio being recorded and data from the design activities being saved for note-taking and analytical purposes. 


\bibliography{ada}

\appendix
\section{Method Resources}               
\subsection{Interview Guide}
\label{appendix:interviewguide}
Below we offer an excerpt of the guide used in the semi-structured interviews. The questions were only accessible by team members and not shared with participants beforehand.

\textbf{PART 1: Notions of “Meaningful Documentation”}

\begin{itemize}
    \item To start, I wanted to briefly learn a bit more about your background. You had mentioned in the interest form that you do work around [insert based on interest form]. Very briefly, in a couple of sentences, can you elaborate on [insert question tailored to interest form response]?
    \item I’m hoping to learn about the approaches you currently take to learn about AI systems in your work. Is there a prior experience in which you were learning about or investigating a harm that was created by an AI system deployed by a technology company? Can you briefly elaborate on the experience in 1-2 sentences.  
    \begin{description}
        \item [Follow up:] Can you elaborate on the strategy or approach you took, to learn this information?  
        \item  [Follow up:] What methods were used?  
        \item  [Follow up:] Did you use any particular artifacts, to make sense of how this harm arose from the algorithmic system?  
    \end{description}

    \item Was there any information about the AI system that you wanted to learn, but didn’t have access to? 
    \begin{description}
        \item [Follow up:] What would have made it easier for you to access the information you needed?
    \end{description}
    \item Imagine you had a magic paper that gave you answers to any questions you asked about how an AI system worked. This could include how the technology company ideated, created, evaluated, or used a given AI system. I’m going to ask you to brainstorm which questions you think are important to disclose to folks in your field, to support work around anticipating/mitigating harms from AI systems or keeping companies more accountable for their AI systems. As you brainstorm these questions, I’ll write them down on these post-its on the shared online board. I’ll check back with you periodically to make sure I’m representing your thoughts accurately. \textbf{So, what critical questions or information types would you want to ask the technology company about, to disclose more meaningful information about the AI system you are investigating?}
\end{itemize}

\textbf{PART 2: Artifact-guided reflection and design}

\begin{itemize}
    \item Imagine you were given this documented information about the system you were studying. Would you find this helpful?
    \begin{description}
        \item [Follow Up:] Have you used these types of artifacts in your own work before?  
        \item [Follow Up:] Are there alternative artifacts that you currently use?
    \end{description}
    \item Given these questions you raised earlier [in part 1], which parts of this artifact would you want to change or improve? 
\end{itemize}

\textbf{PART 3: Ideating alternative artifacts and processes}

\begin{itemize}
    \item We’re on the last set of questions now. So far, we’ve looked at existing artifacts that technology companies use to help govern their AI development practices. Upcoming regulation may prompt companies to disclose the information they document in these artifacts, to better promote collaboration and accountability across stakeholders. Based on the artifacts you saw in this study, do you think existing documentation efforts are helpful for this purpose? Why/why not?  
    \item Imagine that today, the government created new regulation requiring all technology companies to disclose information in response to these questions on this online board. Yet, time-travelling a few years into the future, we still see similar types and degrees of societal harms resulting from industry-created AI systems. What do you envision went wrong?  
    \item Now, imagine that there is a magic genie, with very specific interests. Their goal is to grant wishes that improve technology governance practices, by promoting more complementary and collaborative efforts between civil society and the tech industry. What would you wish for?  
    \item Knowing that it is impossible to create perfect AI, and some harms may be inevitable, what would you consider a “success” for AI governance? For example, imagine that a few decades from now, you are retired. You reflect back on the work you and your community did to support responsible AI governance practices. What type of end-states would you consider a “success”? 
\end{itemize}




\end{document}